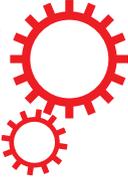

# Naked singularity, firewall, and Hawking radiation

Hongsheng Zhang[1,2,3]

Spacetime singularity has always been of interest since the proof of the Penrose-Hawking singularity theorem. Naked singularity naturally emerges from reasonable initial conditions in the collapsing process. A recent interesting approach in black hole information problem implies that we need a firewall to break the surplus entanglements among the Hawking photons. Classically, the firewall becomes a naked singularity. We find some vacuum analytical solutions in $R^n$-gravity of the firewall-type and use these solutions as concrete models to study the naked singularities. By using standard quantum theory, we investigate the Hawking radiation emitted from the black holes with naked singularities. Here we show that the singularity itself does not destroy information. A unitary quantum theory works well around a firewall-type singularity. We discuss the validity of our result in general relativity. Further our result demonstrates that the temperature of the Hawking radiation still can be expressed in the form of the surface gravity divided by $2\pi$. This indicates that a naked singularity may not compromise the Hakwing evaporation process.

Recently, we hear the incredible gravitational wave signal of a binary black hole merger after a century's wait[1]. Black holes may be the most thoroughly studied objects without being directly observed. Generally a black hole has a singularity enclosed by a horizon (or horizons). According to the Penrose-Hawking singularity theorem, a singularity is inevitable under some general conditions[2,3]. But this theorem does not predict whether the singularity is enclosed by a horizon. If a singularity is not enclosed by a horizon such that an observer at infinity region can see it, it is called a naked singularity. A naked singularity leads to several bad properties of the host spacetime, such as stability problem, causal problem etc. On the other hand it is also helpful to explain some inexplicable astronomical phenomena and to study fine structure of the spacetime in the lab if it really exists. In fact, numerical calculations hint that a naked singularity may appear in gravitational collapse of dust[4–6]. Furthermore, it is shown that gravitational collapse of a scalar field leads to a naked singularity with reasonable initial conditions[7,8]. Even in the frame of loop quantum gravity, a naked singularity could exist, though its physical meaning is slightly different from a classical one[9].

A recent dramatic progress in the studies of black hole information, i.e., the firewall paradox, indicates that naked singularities may be more widespread than what we thought before[10,11]. Essentially, the paradox says that for a black hole the assumption that a free falling observer senses nothing special when crossing the event horizon contradicts a unitary quantum theory. Actually, this assumption is just the equivalence principle, which is the foundation of general relativity. Also, there is no evidence of unitarity violation in nature. General relativity and unitary quantum theory sharply conflict with each other here. According to Almheiri-Marolf-Polchinski-Sully's suggestion, the most economic solution to this contradiction is to assume that there is a firewall with infinite energy density located at the horizon. So the starting point is that one sees nothing at the horizon, while the conclusion is that she will see a firewall. That's really a paradox. Interestingly, a recent study shows that one may obtain the information of the structure of a black hole firewall from the observations of gravitational waves[12].

For the sake of later demonstration, we would like to be involved in a little more details of this paradox. If both the collapse and evaporation of a black hole are unitary processes, the Hawking radiation has to be deviated from an exact Planck distribution. We restore data collapsed into the black hole via the entanglement of the Hawking radiation. Because of the monogamy of quantum entanglement, only two photons can be entangled if they are maximally entangled. No third particle can be involved in such an entangled pair. After the hole is completely destroyed through Hawking evaporation, the whole mass of it will be transferred into the Hawking radiation. The Hawking radiation is in a pure state, full of entangled pairs (if the progenitor of the black hole is in a pure state).

[1]School of Physics and Technology, University of Jinan, West Road of Nan Xinzhuang 336, Jinan, Shandong, 250022, China. [2]Center for Astrophysics, Shanghai Normal University, 100 Guilin Road, Shanghai, 200234, China. [3]State Key Laboratory of Theoretical Physics, Institute of Theoretical Physics, Chinese Academy of Sciences, Beijing, 100190, China. Correspondence and requests for materials should be addressed to H.Z. (email: hongsheng@shnu.edu.cn)

Here we present a simple explanation of the Page time. The Page time is the time when the von Neumann entropy of the Hawking radiation outside a black hole reaches a maximum[13]. Assume the total number of photons emitted from the black hole in its lifetime is $N$. The $N/2$-photons emitted in the early stage of the black hole (called young hereafter) are entangled with the other $N/2$-photons emitted in the late stage (called old hereafter). For a young black hole, the entropy of the surrounding Hawking radiation increases, since an observer outside the black hole does not know that a photon outside the horizon has a spouse hidden in the black hole. For an old black hole, every photon becomes to find its spouse one by one, and therefore the total entropy of the Hawking radiation decreases. In spite of increasing or decreasing of the entropy, the temperature always increases in the evaporation process. According to statistical physics, the average energy of a photon $\epsilon_a$ is proportional to the temperature $T$. The temperature $T$ is proportional to $1/M$, where $M$ is the mass of the black hole. We consider the case for $M \gg M_P$, where $M_P$ denotes the Planck mass. The behavior of a black hole with mass $M \sim M_P$ is still controversial. Maybe only a full quantum gravity theory can deal with it. Supposing that the number of the total Hawking photons reaches $N/2$ when the the black hole mass shrinks to $M_h$, we have

$$\frac{N}{2} = \int_{M_P}^{M_h} \frac{dM}{\varepsilon_a} = \int_{M_h}^{M_i} \frac{dM}{\varepsilon_a}, \qquad (1)$$

where $M_i$ denotes the initial mass of the black hole. It is easy to obtain the Page mass corresponding to the Page time $M_h = M_i/\sqrt{2}$. In fact, a black hole also emits gravitons and massive particles. And thus the realistic Page time will be a little smaller than what we have shown above. A detailed numerical analysis confirms this point[14].

For an old black hole, every Hawking photon emitted has its spouse in the thermal bath outside the black hole. Thus, it cannot be entangled with anything inside the black hole. Mathematically, this indicates that a field configuration cannot be continuous when crossing the horizon[11,15]. So the kinetic term, and thus the Hamiltonian of the field and pressure diverge at the horizon. Classically, a scalar polynomial curvature singularity appears in the frame of general relativity. Generally, a divergent matter density leads to a divergent scalar polynomial of curvature in general relativity. One can demonstrate this point as follows. For a sourced spacetime the Ricci scalar $R$ and a scalar polynomial of curvature $G_{\mu\nu}G^{\mu\nu}$ are related to the density $\rho = -T_0^0$ and pressure $p_i = T_i^i$ (not sum over $i$) of the fluid by $R \sim -(\rho + \Sigma_{i=1}^3 p_i)$, and $G_{\mu\nu}G^{\mu\nu} \sim (\rho^2 + \Sigma_{i=1}^3 p_i^2)$, respectively. Here $T_{\mu\nu}$ denote the stress energy tensor. Proper $p_i$ and $\rho$ may control the divergence of $R$, but a divergent $G_{\mu\nu}G^{\mu\nu}$ always follows a divergent density $\rho$ and $p_i$. If the putative firewall does exist, the naked singularities may be more widely distributed in the universe than what we expected before. We emphasize that the Page time for a solar mass black hole is about $10^{72}$ seconds, which is much longer than the age of the universe. For a black hole whose mass is about $10^{15}$g (primordial black hole), the Page time is approximately equal to the age of the universe (However, it is still controversial about the epoch when the firewall comes into being. If the "fast scrambling" argument is taken into account, the firewall will be formed in 0.1 seconds for a solar mass black hole[10,14,16]). One may try to find some classical solution to simulate the firewall through considering its back reaction to the original spacetime. Unfortunately we know little about the stress-energy and other physical properties of the firewall except its divergent density and pressure. Furthermore, to consider the back reaction of a divergent physical quantity has its inherent difficulty, i.e., a very large quantity cannot be treated as back reaction.

Note that it is suggested in ref. 15 that the curvature singularity at $r=0$ shifts out to the horizon to form a firewall. And the spacetime behind a firewall does not exist. Thus, it cannot be called a black hole in a rigorous sense, since there is no horizon enclosing such a firewall singularity. In this paper, we adopt a more conservative point. That is, the spacetime behind the firewall survives after the formation of the firewall wall[10,11,14,17]. The region enclosed by the horizon is still a black hole in the classical sense. No particle can escape from this region.

A black hole with scalar polynomial curvature singularity at the horizon should be treated as an appropriate classical toy model of the black hole firewall required by quantum theory. Because the "realistic" naked singularities in general relativity[4–8] are rather complicated and have no analytical forms, we turn to some toy models with naked singularities in modified gravity. Previous experiences show that proper toy models are very conducive to the studies of realistic cases. Up to now we know little about the quantum mechanics around the naked singularities in spite of significant progresses in quantum mechanics in curved spaces. Hawking radiation takes a pivotal status in black hole physics. One may expect that the effects of quantum mechanics, especially the Hawking radiation, will be significantly affected by the naked singularity. Now we study the Hawking radiation of spacetimes with naked singularities.

Before investigating a concrete metric, we consider a generic static metric as follows,

$$ds^2 = -f(r)dt^2 + h^{-1}(r)dr^2 + r^2 d\Omega_2^2, \qquad (2)$$

where $d\Omega_2^2$ is a maximally symmetric two dimensional space. In 1974 Hawking showed that the vacuum state of a scalar field near the horizon is different from the vacuum state at the asymptotically flat infinity region for a collapsing Schwarzschild black hole[18]. By using the Bogoliubov transformation between different vacuums, one finds that an initial observer at spacelike infinity senses an exact thermal spectrum. The temperature of the Hawking radiation is $\kappa/2\pi$, where $\kappa$ denotes the surface gravity at the event horizon, which coincides with the temperature obtained from black hole mechanics by analogy[18,19].

Several different methods have been proposed after Hawking's original work, including thermal field theory method, topological method, quantum anomaly method, and a few others[20]. All of these methods lead to the same temperature for a Schwarzschild black hole. However we shall see that they do not match each other for the general case (2). According to its original definition, the surface gravity $\kappa$ is the magnitude of the 4-acceleration for a particle resting at the static coordinates multiplied by a red shift factor $\sqrt{f}$, $\kappa = \frac{1}{2} f' \sqrt{\frac{h}{f}}$, where a prime denotes a derivative with respective to $r$. The corresponding temperature reads, $T_s = \frac{1}{4\pi} f' \sqrt{\frac{h}{f}} \Big|_{r=r_h}$. Here $r_h$ is the radius of the horizon. On the other hand thermal field theory presents, $T_t = \frac{1}{4\pi} \sqrt{h'f'} \Big|_{r=r_h}$. It is easy to see that $T_s = T_t$ if $h = f$. But when $h \neq f$, the two temperatures match each other only if, $\frac{f'}{f}\Big|_{r=r_h} = \frac{h'}{h}\Big|_{r=r_h}$. Now we consider solutions in modified gravity with naked singularities, which break the condition $h = f$. These solutions also can be realized by introducing some matter fields in general relativity. For simplicity we first study a topological black hole in $R^n$-gravity. For detailed field equation for the $R^n$-gravity, see for example ref. [21]. We find the following solution for $R^n$-gravity,

$$f = \left(1 - \frac{M}{r}\right)^3, \quad h = \frac{16384}{19683} \frac{M\left(1 - \frac{M}{4r}\right)^7}{r - M}. \tag{3}$$

Here $M$ is a mass parameter. And $d\Omega^2 = dx^2 + dy^2$. Thus, it is a topological black hole. In this spacetime, we have $T_s = \frac{1}{4\pi M}$, and $T_t = \frac{i}{4\sqrt{3}\pi M}$. Clearly, $T_t$ does not indicate the correct answer. One can show that the singularity at the horizon $r = M$ is a scalar polynomial curvature singularity in the sense $W^{\alpha\beta\gamma\delta}W_{\alpha\beta\gamma\delta} = \frac{4M^2(M-4r)^{12}(5M^2 - 10Mr + 2r^2)^2}{129140163(M-r)^4 r^{18}}$. Here $W$ denotes the Weyl tensor. Just as the case of firewall, this singularity is "critically" naked. On one hand, there is no horizon enclosing this singularity, thus it is naked. On the other hand, a particle emitted from the singularity will be infinitely redshifted at infinity. So an observer at infinity cannot see it directly. In this sense, it is not naked. But if we consider the quantum effects, the singularity will be stretched to some finite thickness (stretched horizon). Then this singularity can really be seen by an exterior observer. Fortunately singularity is not so formidable as one usually thinks. We shall see that it is tractable in the quantum treatment.

Next, we derive the Hawking radiation via Damour-Ruffini's method[22], which does not depend on a collapsing black hole. We consider a massless scalar field $\phi$ in the above black hole background,

$$\frac{1}{\sqrt{-g}} \frac{\partial}{\partial x^\mu} \left(\sqrt{-g} g^{\mu\nu} \frac{\partial \phi}{\partial x^\nu}\right) = 0. \tag{4}$$

To separate variables, we set $\phi = e^{-i\omega t} \varphi(r) \psi(x, y)$. We assume $\psi$ to be a harmonic function of $x$ and $y$. Thus $\phi$ becomes a two dimensional wave function. The radius equation is the key equation after variable separation,

$$h\varphi'' + \left[\frac{2h}{r} + \sqrt{hf^{-1}}(\sqrt{hf})'\right]\varphi' = -\frac{\omega^2}{f}\varphi. \tag{5}$$

We introduce the generalized tortoise coordinate $r_* = \int (fh)^{-1/2} dr$. $r_*$ can be expressed in inverse hyperbolic tangent functions. The most important property of the Hawking radiation depends on the behavior of the wave equation around the horizon. So we expand $r_*$ at $r = M$,

$$r_* = r + 3M \ln \left|\frac{r-M}{M}\right| - \frac{1}{M}(r-M)^2 + \mathcal{O}(r-M)^2. \tag{6}$$

Around the horizon $r = M$, the radius equation becomes $\frac{d^2\varphi}{dr_*^2} + \omega^2 \varphi^2 = 0$. The outgoing wave solution is $\phi_{\text{oute}} = e^{-i\omega v} e^{i2\omega r_*}$, where $v$ is the advanced Eddington coordinate, $v = t + r_*$. Replacing $r_*$ by $r$, we have $\phi_{\text{oute}} = e^{-i\omega v}(r - M)^{i6M\omega}$, which is not well-defined at $r = M$. It is not only a coordinate singularity, but also a scalar polynomial curvature singularity. We make an analytic continuation to interior of the horizon on a conical surface of a complex $r - M$ (that is, endowed with an imaginary part) by an angle $\pi/3$. On this conical surface, to rotate by an angle $2\pi/3$ around the original point means to get back to the starting point. Thus to rotate $\pi/3$ means to reach the antipodal point, which is different from an ordinary complex plane. This is the key technical difference from what we done in the case of an ordinary black hole without naked singularity. The true singularity at the horizon is translated to a conical singularity in the analytical continuation. After this analytical continuation, the outgoing wave is extended to the interior of the horizon, $\phi_{\text{outi}} = e^{-i\omega v}(r - M)^{i6M\omega} e^{2\pi M\omega}$. Comparing to $\phi_{\text{oute}}$, we obtain the scattering factor across the horizon $\Psi = e^{-2\pi M\omega}$. Thus the scattering probability $|\Psi|^2 = e^{-4\pi M\omega}$. By using Sannan's approach[23], we derive the spectrum of outgoing wave $P = \frac{\Gamma}{e^{\omega/T} \pm 1}$, where $T = \frac{1}{4\pi M}$. $\pm$ correspond to Fermion and Boson, respectively, and $\Gamma$ is the factor of barrier penetration for the horizon. One can check that $T_s$ presents the proper temperature. This means that the method of surface gravity is fairly robust. It can give the proper temperature even when a singularity at the horizon appears. On the contrary the method of thermal field theory is defective for this metric.

Now we discuss how to embed the above results into general relativity. First of all, the metric (2) does not depend on any gravity theory. The point is that it yields different stress energy tensors in different gravity theories. For the vacuum solution (3) in $R^n$-gravity, its corresponding stress energy in general relativity reads,

$$T_e = \frac{1}{8\pi}\Bigg(-\frac{M^2(4r-M)^6(r-2M)}{6561(r-M)^5 r^6}dtdt$$
$$+\frac{M^2(4r-M)^{14}(2M+r)}{387420489(r-M)^3 r^{16}}drdr$$
$$+\frac{M(4r-M)^6(4M^2-5Mr-2r^2)}{19683(r-M)^2 r^{11}}d\theta d\theta$$
$$+\frac{M(4r-M)^6(4M^2-5Mr-2r^2)}{19683(r-M)^2 r^{11}}d\phi d\phi\Bigg).$$
(7)

One can check that both the density and pressure corresponding to this stress energy are divergent at the horizon $r=M$, which leads to the curvature singularity at the horizon. This stress energy presents a concrete example for the above general discussions about divergent density and pressure at the horizon. Actually, the calculation of the Hawking radiation only depends on the metric form. All the calculations have no relation to the form of the stress-energy once the metric is given. It always leads to the same result for the same metric. In this sense, we can say that our demonstration of Hawking radiation of spacetime with naked singularity is, in fact, gravity theory-free.

The critical problem of black information, and furthermore quantum gravity, is that whether a singularity leads to a non-unitary evolution. From the above discussions, one sees that an outgoing wave always propagates unitarily when it is scattered by the horizon, where a singularity appears. So a preliminary conclusion is that the unitarity of quantum theory does not compromise in a spacetime with naked singularity. We obtain a black body spectrum since we work on a static background geometry. By introducing the back reaction of the Hawking radiation to the background geometry, one can obtain a modified spectrum, which is implicitly consistent with the unitary theory[24]. One may be puzzled by the operation to circumvent a scalar polynomial curvature singularity. We present some discussions on this problem in order.

(i). First of all the appearance of singularity only implies that the classical gravity fails at that point. Any realistic measurable variable must keep finite. At present stage we believe that an "infinite variable" in gravity theory is in fact at the Planck scale.
(ii). A scalar polynomial curvature singularity is not equal to the singularity at which the geodesics can not be extended[25]. The geodesics still may be extended across a scalar polynomial curvature singularity.
(iii). An extension across a true singularity is not unacceptable in mathematical treatment. For example we make analytical extension to the region $r<0$ in the Kerr space when we plot the corresponding Penrose diagram. In fact the true singularity at $r=0$ for the Schwarzschild spacetime also can be analytically extended[26].
(iv). The scalar polynomial curvature invariants are not divergent on the path of the analytical continuation. Our calculations are proceeded in the well-defined space all the time.
(v). In brane world models, the bulk spacetime has a scalar polynomial singularity where the brane dwells. The gravitons propagate across the brane without any divergence.
(vi). As a very important point, this scalar polynomial curvature singularity at the horizon may be a proper classical simulation of the black hole firewall. There is no firewall for a classical black hole. It appears when one considers quantum theory. It can be treated as a quantum anomaly. But this anomaly is so large that a spacetime singularity appears at the horizon. The firewall makes an infalling observer burn up at the horizon to break quantum entanglement with anything outside the hole. So its affection an outgoing Hawking radiation deserves exploration. We demonstrate that a black hole firewall does not compromise the Hawking radiation. The Hawking radiation propagates as usual in a spacetime in which a firewall-type singularity appears (Fig. 1).

In the above context, we discussed a topological black hole. Now we consider a more realistic example, that is, a spherical symmetric black hole. For this black hole in $R^n$-gravity, $f=-(1-M/r)^2$, $r\geq M$; $f=(1-M/r)^2$, $r<M$, and $h=1$. Note that this solution is not a limited form of the Reissner-Nordström black hole. There is a scalar polynomial curvature singularity at $r=M$ since $W^{\alpha\beta\gamma\delta}W_{\alpha\beta\gamma\delta}=\frac{12M^2}{(r-M)^2 r^4}$. Mimicking the previous calculations, we obtain the temperature of the Hawking radiation $T=\frac{1}{2\pi M}$. A special note is that the angle of rotation of $r-M$ is $\pi/2$ on a conical surface to get back to the antipodal point when one makes an analytic continuation to penetrate the horizon. One can confirm that $T_s$ presents the right temperature once again, while the temperature given by thermal field theory is wrong.

For more complex spaces in modified gravity, we have similar conclusions. In a little more complicated example in $R^n$-gravity, $f=\sqrt{1-\frac{4M}{r}}$ for $r\geq 4M$, $f=-\sqrt{\frac{4M}{r}-1}$ for $r\leq 4M$, and $h=2^{-5/7}\frac{49M}{4r}(1-\frac{4M}{r})^{3/2}(1-\frac{7M}{2r})^{-11/7}$ for $r\geq 4M$, $h=2^{-5/7}\frac{49M}{4r}\left(-1+\frac{4M}{r}\right)^{3/2}\left(1-\frac{7M}{2r}\right)^{-11/7}$ for $r\leq 4M$, where $\Omega^2=dx^2+dy^2$. By using the previous procedure, one can demonstrate that the temperature of the Hawking radiation is $T=T_s=\frac{7}{16\pi M}$. In this case, the angle of rotation of $r-4M$ is $2\pi$ on a Riemann surface $z^{1/2}$. A rotation with angle $2\pi$ around the original point does not mean to get back to the starting point on a Riemann surface. One can check that thermal field theory still cannot present correct temperature in this case.

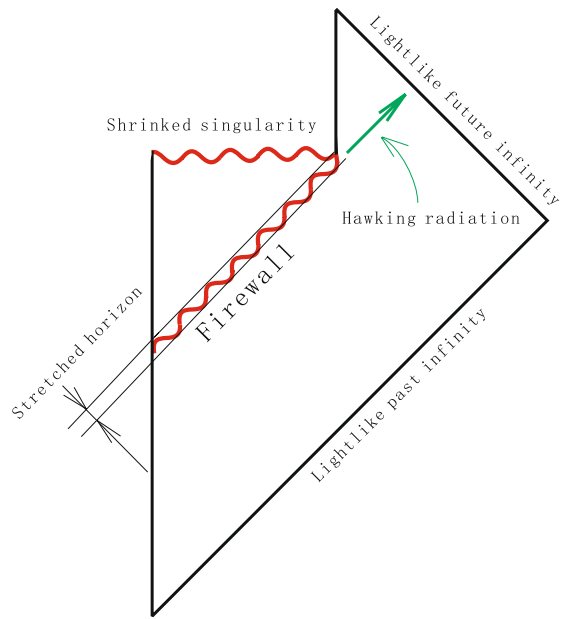

**Figure 1.** The conceptual Penrose diagram for an old black hole. This diagram shows an old black hole. The naked firewall has been formed around the horizon.

In all theories, it is a marvel that general relativity can predict where it fails by itself. This is the singularity. We demonstrate the properties of the Hawking evaporation of the spacetimes with naked singularities. The Hawking photons evolve unitarily when crossing the singularity. This result indicates that a firewall may not lead to a non-unitary evolution of the Hawking photons. The thermodynamics of a naked singularity is far from complete. Now we just have some preliminary results by the aids of the AdS/CFT correspondence[27, 28]. Especially, the Gibbons-Maeda-Garfinkle-Horowitz-Strominger (GMGHS) dilaton black hole has a null singularity at the extremal limit[29, 30], which is similar to a firewall singularity at the horizon. The temperature of the extremal GMGHS black hole is $\kappa/(2\pi)$[31], which coincides with our result directly derived from the Hawking radiation. Our study sheds some light on the quantum effects for naked singularities and lays a general foundation for a sound thermodynamics for spacetimes with naked singularities. Einstein and Rosen thought that the horizon of the Schwarzschild is a true singularity, and thus should be removed from realistic spacetimes[32]. It is believed that a singularity will ruin all physical laws around it. Now we show that the theory may be tractable even when there is a true singularity at the horizon.

### Acknowledgements

This work is supported by the Program for Professor of Special Appointment (Eastern Scholar) at Shanghai Institutions of Higher Learning, National Education Foundation of China under grant No. 200931271104, and National Natural Science Foundation of China under Grant Nos 11075106, 11575083 and 11275128.

### Author Contributions

H.Z. thanks J. Sully, L. Li, X. Wu and Y. Ong for helpful discussions.

### Additional Information

**Competing Interests:** The authors declare that they have no competing interests.

**Publisher's note:** Springer Nature remains neutral with regard to jurisdictional claims in published maps and institutional affiliations.